# Large Area Roller Embossing of Multilayered Ceramic Green Composites

Xuechuan Shan[+], Y. C. Soh, C. W. P. Shi, C. K. Tay and C. W. Lu

Singapore Institute of Manufacturing Technology (SIMTech),
71 Nanyang Drive, Singapore 638075

[+] Corresponding Email: xcshan@simtech.a-star.edu.sg (Xuechuan Shan)

*Abstract* -This paper presents our latest achievements in developing large area patterning of multilayered ceramic green composites using micro roller embossing. The aim of this research is to develop large area pattern technique for ceramic green substrates using a modified roller laminator, which is compatible with screen printing apparatus, for integration of micro embossing and printing in the future stage. A thin film nickel mold was developed via photolithography, nickel electroplating and photoresist strip-off. The mold had an effective panel size of 150 mm× 150 mm with the height of protrusive micro patterns being about 40 μm. Formation of micro patterns was successfully demonstrated over the whole panel area using roller embossing on laminated ceramic green tapes (HL2000 from Heraeus). Micro patterns for inductors, capacitors as well as interconnection with 50 μm line width were embossed on ceramic green substrates. With the optimized process parameters (including feeding speed, roller temperature and applied pressure), we have demonstrated that micro roller embossing is a promising method for large area patterning of ceramic green substrates.

## I. INTRODUCTION

Ceramic materials such as Low temperature cofired ceramics (LTCC) have been used for many years in microelectronics industries and is currently in use for a variety of emerging applications such as microfluidic devices, bio-medical devices, microsystems and three dimensional modules for telecommunication devices [1]. The conventional processes for LTCC fabrication include patterning vias and cavities on individual ceramic green tapes, printing on surfaces, filling vias, lamination as well as debinding and sintering [2, 3]. The patterns such as micro cavities, channels and vias were conventionally fabricated via mechanical punching [3] or laser drilling [4] of ceramic green tapes. These conventional methods have critical limitations, for instance, the minimum depth to be machined is limited by the thickness of a green tape; and the lateral dimensional of a minimum feature is limited by the size of machining tools or diameter of laser beams. Formation of noncircular micro patterns such like micro channels or tapered features is more challenging for mechanical punching or laser drilling.

One of the alternative methods to generate micro patterns on ceramic green substrates is micro embossing, which has been widely used for generating micro structures on polymers or glasses for microfluidic and bio-medical devices [5-7]. The use of micro embossing for patterning green ceramic tapes will enhance the advantages of LTCC as a substrate material, and open new fields of emerging applications for this material. The researchers in Technical University of Berlin have demonstrated the feasibility of patterning ceramic green tapes using simultaneous embossing, i.e. embossing with a flat mold and heater [8]; a group in Vienna University of Technology has demonstrated micro wedges patterned on ceramic green substrates using simultaneous embossing [9].

A green ceramic tape is composed of ceramic-based powders and polymeric additives, which can be categorized as binders and plasticizers in terms of their functions. Due to its ingredient, the green tape or laminated green substrate is soft and flexible with low strength; this flexibility and low strength will increase the difficult in demolding, since an embossed green substrate could be possibly torn up or seriously warped during demolding.

In this study, we proposed micro roller embossing [10, 11] for generating patterns on large area ceramic green substrates. The final aim is to integrate micro embossing and screen printing in future stage. Comparing with the popular simultaneous embossing, the advantages of roller embossing are: (1) localized contact area with heater so that only small area of substrate is headed; (2) localized embossing area that improves easiness in demolding; (3) lower embossing force for a desired embossing process; and etc. A thin nickel mold was developed via nickel electroplating 40 μm protrusive patterns on a nickel film of 75 μm thick. The thin film mold has an effective panel size of 150 mm× 150 mm, which is one of the standard panel sizes for screen printing. Formation of micro patterns was successfully demonstrated using roller embossing on laminated ceramic green tapes (HL2000 from Heraeus) over the whole panel area. Micro patterns such as inductors and capacitors were formed on the ceramic green substrates. The achievements demonstrated that micro roller embossing is a promising method for patterning large area ceramic green substrates.

## II. EXPERIMENTAL METHODOLOGY

### A. Micro Roller Embosser

The micro roller embosser is a modified version of general purpose thermal laminator, which consists of two rollers, i.e. the upper roller 1 and bottom rollers 2, as shown in Fig.1. The upper roller 1 is an active roller and made of steel with an embedded heater; the temperature of which can be preset up to 200 °C. The bottom roller is a passive one made of hard rubber without the embedded heater. The diameter of both rollers is 100 mm with the width being 400 mm. The pressure for embossing and rotation for substrate feeding can be pre-applied to roller 1 via a controller. The maximum pressure after modification can be set up to 14 bars, and the feeding speed ranges from 0 to 20 mm/sec. The surface temperature over the roller width was measured and its stability and uniformity characterized by means of a portable temperature sensor. It was found that the surface temperature actually measured via the sensor was quite different from the setting temperature preset via the controller. However, the variation of temperature measured over the roller surface, as shown in Fig. 2, was within ± 2 °C, which was well acceptable.

### B. Thin Film Nickel Mold

A nickel thin film with a thickness of 75 µm was used as the substrate for mold making. The nickel film, the overall panel size of which was 200 mm× 180 mm, was coated with a layer of photoresist, followed by UV patterning, nickel electroplating and photoresist strip-off. By precisely controlling the parameters for electroplating (solution concentration, temperature, PH value and electric current), a nickel film mold with an effective panel size of 150 mm× 150 mm was obtained. The height of electroplated protrusive patterns was approximately 40 µm and the total thickness of the film mold was 115 µm, which was flexible enough to be wrapped on the roller if necessary.

There were 6 identical units over the effective area of the film mold, as shown in Fig.3, the upper three units are named from left to right as $U_L$, $U_C$ and $U_R$, and the lower three units as $L_L$, $L_C$ and $L_R$, respectively. Fig. 4 shows one unit $U_C$ among them. Two groups of patterns (group A and group B) in each unit were targeted for characterization to explore the pattern quality. Fig. 5 shows the 75 µm line/ 300 µm pitch in Group A, it can be seen that the patterns were about 40 µm high, and uniformly distributed on the film mold. Fig.6 illustrates a three dimensional measurement of a square inductor in Group B on the nickel mold.

Instead of wrapping the film mold on roller 1, an alternative solution, as illustrated in Fig. 7, was used for roller embossing, i.e. the film mold was aligned and then stacked with the ceramic green substrate and supporting plate in a sandwich structure. The supporting plate was made of FR4 board and was used to provide a hard surface for roller embossing to minimize the influence of the rubber roller. Comparing with the approach of wrapping the film mold directly on the roller, our method illustrated in Fig. 6 has following advantages of: (1) easiness of mold attachment or replacement; (2) less limitation in mold thickness and lateral dimension.

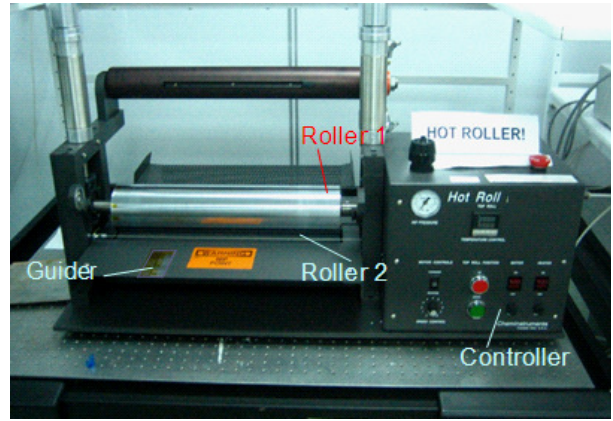

Fig. 1. The micro roller embosser used in this study. Roller 1 was a driving roller with an embedded heater inside; and roller 2 is a supporting roller made of hardened rubber.

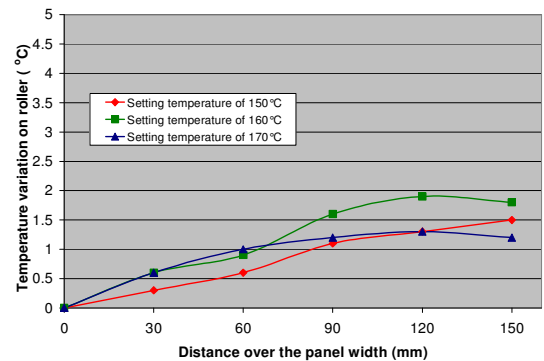

Fig. 2. Variations of temperature on roller surface over the panel width. The measurement started from the guider and the temperature slightly increased from left to right along the roller width.

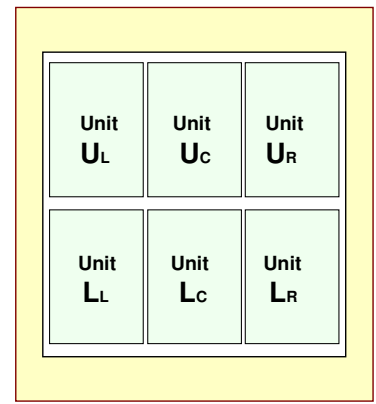

Fig. 3. The six identical pattern units designed on the film mold

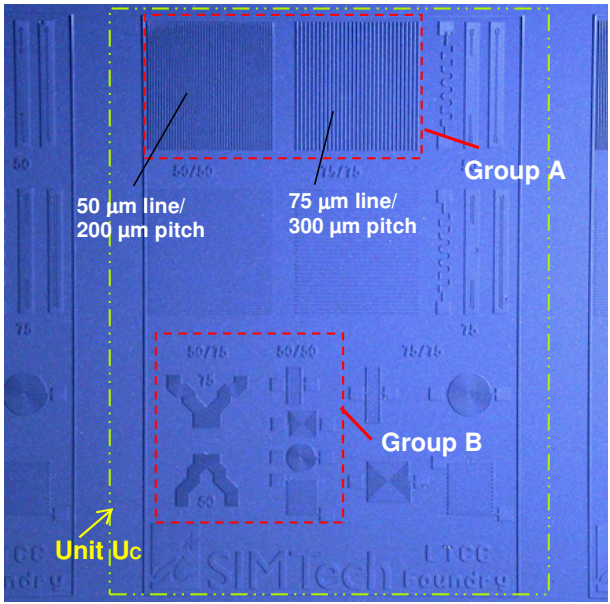

Fig. 4. One of the six units on the electroplated nickel film mold

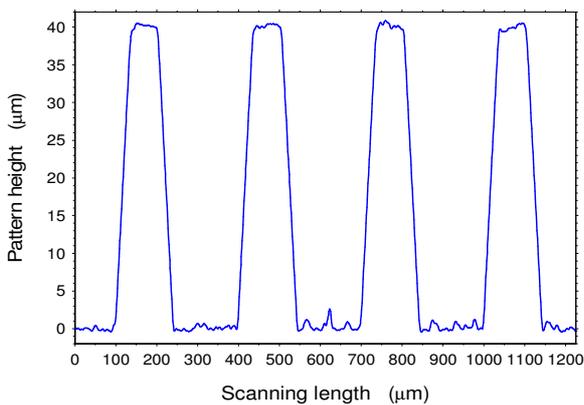

Fig. 5. Pattern profiles of the 75 µm line/ 300 µm pitch on the mold

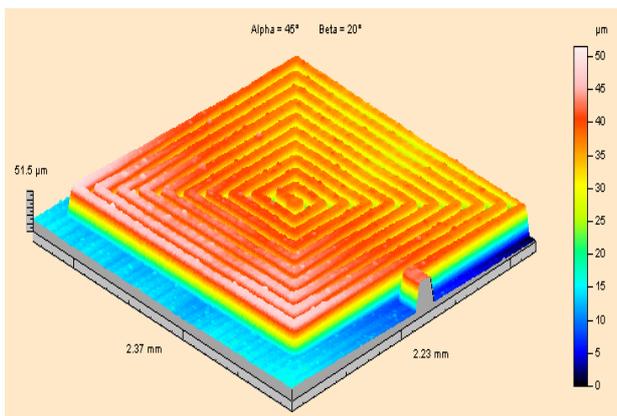

Fig. 6. Three dimensional measurement of a square inductor in Group B on the nickel mold. The height of protrusive patterns was 40 µm.

## C. Ceramic Green Substrates

The micro roller embossing started from laminating ceramic green tapes to a stacked substrate. Commercially available low temperature co-fired ceramic (LTCC) green tapes with a brand name of HL2000 from Heraeus, Germany were used in this study, which has a low shrinkage ratio in its x-y plane and high shrinkage ratio in its thickness direction. By using an isostatic laminator with lamination pressure, temperature and holding time being 10 MPa, 70 °C and 10 minutes, respectively, four layers of such green tapes (150 mm× 150 mm) were laminated to a 0.5 mm thick substrate for roller embossing.

### III. EXPERIMENT RESULTS

### A. Roller Embossing Process

Micro roller embossing started from presetting of process parameters (roller pressure, temperature and feeding speed) via its controller. After parameter setting and then stabilization for about 10 minutes, the film mold, the ceramic green substrate and a FR4 supporting plate were sandwich structured and aligned with the guider for feeding. The driving roller, i.e. roller 1, was then lowered down to press the sample with a constant pressure. As the rotation of roller 1 was driving the sample to move through the roller-pair with a constant speed without slip, micro patterns were then replicated on the green substrate with reversed profiles.

The ratio of polymeric additives versus ceramic powders was only 16.5 wt% for HL2000 green tapes. This low ratio would slow down the material flow during roller embossing even at raised temperature, and the pattern formation in ceramic embossing is supposed to be a combined effect of plastic deformation and material flow. Nevertheless, roller embossing on large area ceramic green substrate has been successfully demonstrated. Fig. 8 shows micro patterns on a 150 mm× 150 mm substrate with preset parameters of 14 bars, 160 ºC and 1.6mm/sec. Figs. 8 (a) and (b) show the conductive patterns on the film nickel mold and that embossed on a ceramic green substrate; Figs. 8 (c) and (d) show a square inductor on the mold and the pattern embossed on the green substrate, respectively. The micro patterns on the nickel mold were replicated on the substrate with a reversed profile.

### B. Process Uniformity

In order to explore the uniformity within the effective panel size of 150 mm× 150 mm, the same pattern with a step-shaped profile on each of six units was selected and characterized to evaluate the process uniformity. Fig. 9 demonstrates variations of the embossed depth of the selected pattern at 160 °C, 14 bars and 1.6 mm/sec. it was found that the lowest embossed depths were observed in the center pattern regions (with the legend of $L_C$ and $U_C$ in Fig.3). This may be due to the uneven pressure applied on the mold-substrate pair, the uneven pressure was possibly caused by the following reasons: (1) the diameter tolerance or straightness of rollers, or (2) the air bubbles trapped during embossing.

The variations in embossed depths between the left and right regions ($L_L$, $U_L$ versus $L_R$, $U_R$) were within 2.0 µm. This might be due to the variations of the surface temperature shown in Fig. 2, and uneven pressure distribution as well. More detailed investigations to identify the specific reason are being executed.

In order to explore the uniformity within an embossed pattern, 75 µm lines/ 300 µm pitches in Group A of the nickel mold were selected, and the embossing results were shown in Fig. 10. The applied pressure and feeding speed were 12 bars and 1.6 mm/sec, respectively. The embossed depth and pattern uniformity were plotted and compared as the embossing temperature was set to 160 °C and 170 °C, respectively, via the controller.

From the profiles shown in Fig.10, it was found that the embossed channels were deeper near the edge location than that near central region within the pattern group. This was due to the variation of the localized pressure distribution within the patterns. However, as the temperature increased from 160 °C to 170 °C, this variation decreased and became less obvious. It was also found that protrusive features were higher near the edge locations than that in the central region. This might be due the following reasons: (1) the pull-out effect elongated the features at the edge locations during demolding; (2) the material flow helped green materials filling into the mold cavity. More detailed investigations are needed to identify the reasons and solutions for improving the pattern uniformity.

*C. Effect of Process Parameters*

The effects of main process parameters were investigated. Fig. 11 illustrated the effect of applied pressure versus the embossed depth of channels in group A (75 µm width and 300 µm pitch) at 1.6 mm/sec and 170 °C. It was observed that the embossed depth significantly increases as the applied pressure increased from 8 bars to 14 bars. Fig. 12 illustrated the effect of roller temperature versus the embossing depth of the same channels under the pressure of 10 bars and feeding speed of 1.6 mm/sec. It was observed that the embossed depth significantly increases as the roller temperature increased from 100 °C to 170 °C. Fig. 13 demonstrated the impact of feeding speed against the embossing depth at 150 °C and 10 bars. It was seen that the embossed depth was significantly decreased as the feeding speed increased from 0.67 mm/sec to 9.8 mm/sec.

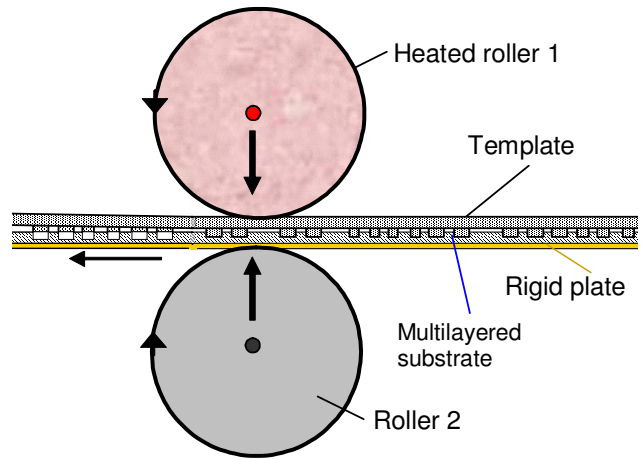

Fig.7. The method of micro embossing adopted in this study. The substrate is sandwiched with the mold and a rigid supporting plate

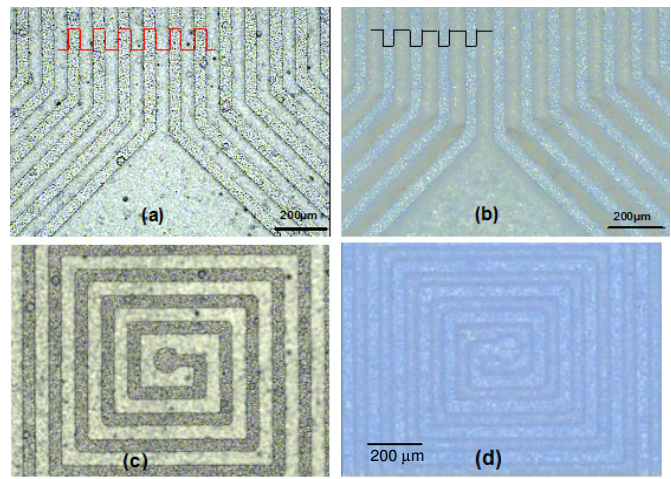

Fig. 8. The conductive patterns (a) on the mold and the conductive patterns (b) embossed on the ceramic green substrate. A square inductor (c) on the mold and the pattern (d) embossed on the green substrate.

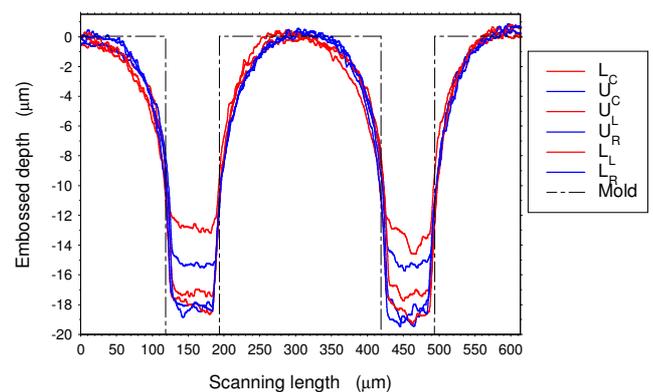

Fig. 9. Variations of embossed depth of the same patterns in six units within the panel at 160 °C, 14 bars and 1.6 mm/sec.

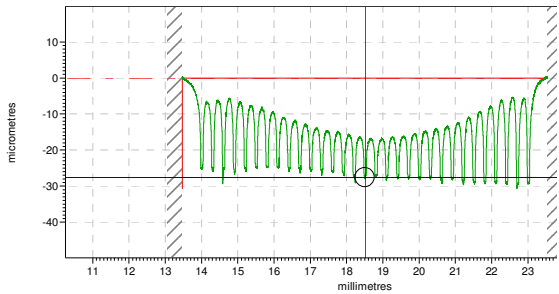

(a) At 160 °C, 12 bars and 1.6 mm/sec.

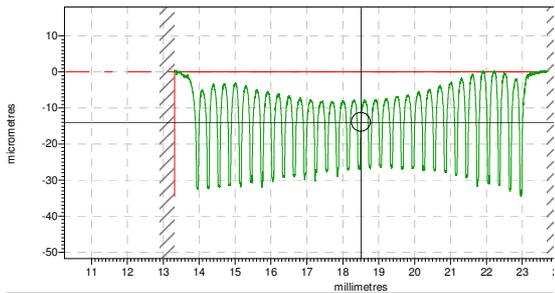

(b) At 170 °C, 12 bars and 1.6 mm/sec

Fig. 10. Variations of the embossed depth versus preset temperatures of the same patterns within pattern unit $U_C$

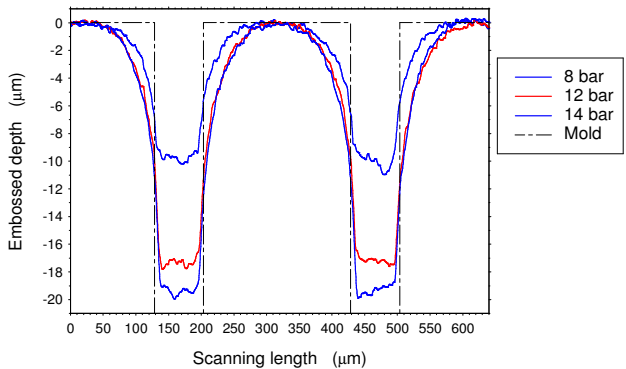

Fig. 11. The effect of pressure versus the embossed depth within pattern unit $L_C$ (170°C and 1.6 mm/sec).

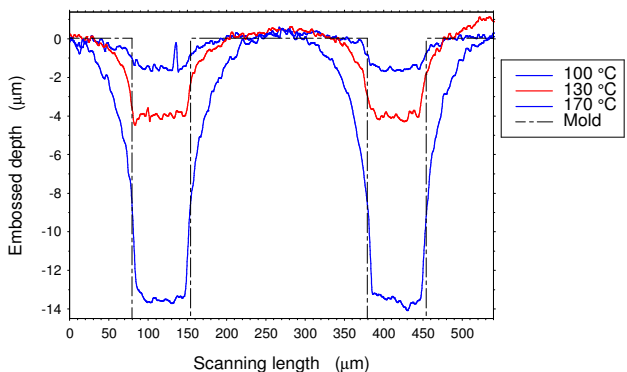

Fig. 12. The effect of temperature versus the embossed depth within $L_C$ (10 bars and 1.6 mm/sec).

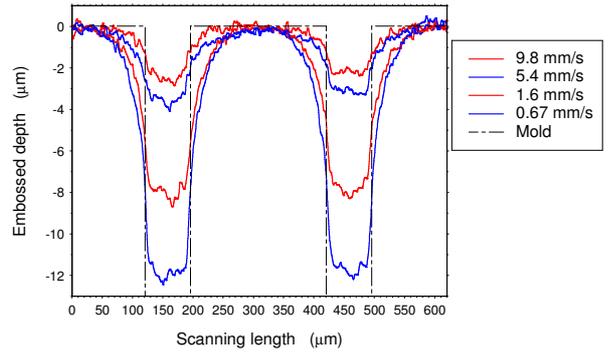

Fig. 13. The effect of feeding speed versus the embossed depth within pattern $U_C$ (150°C and 10 bars)

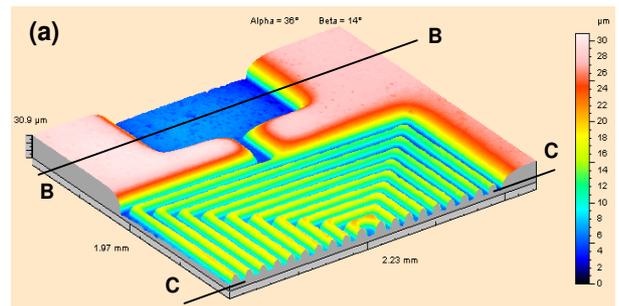

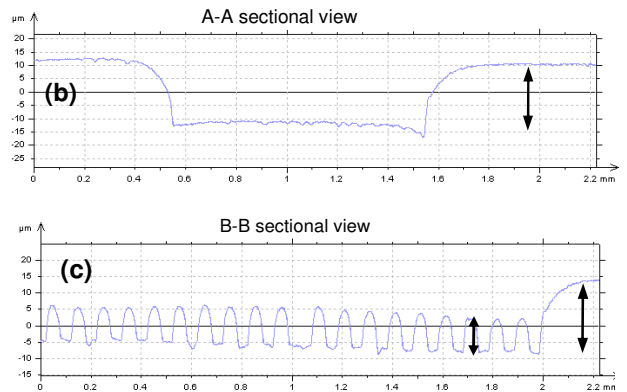

Fig. 14. Stylus results of an embossed inductor pattern in (a) and cross sectional views of the anchor region in (b) and inductor lines in (c)

Fig. 14 (a) demonstrates a three dimensional result of the square inductor pattern embossed on the green substrate as illustrated in Fig. 8, the line/pitch ratio of which was 50 µm/ 100 µm. The cross-sectional view of the anchor region and that of the inductor lines were shown in Figs. 14 (b) and (c), respectively. These results indicated the influence of pattern density against the embossed depth. An overall recessed deformation was observed within the inductor pattern in Fig. 14(c). This phenomenon of overall recessed deformation, which was also observed in another report [12], would occur with the increase of pattern density; and it can be improved by further optimizing the embossing conditions and the pattern design on the mold.

## IV. Conclusions

Micro roller embossing on large area ceramic green substrates was successfully demonstrated by means of a mechanically modified thermal laminator. By using an electroplated film mold and sandwiched mold-substrate-supporting plate structure, formation of micro patterns on ceramic green substrates over an effective panel size of 150 mm× 150 mm was performed; and micro patterns such as inductors, channels with smallest line width of 50 µm were embossed on the ceramic green substrates with good pattern quality. This investigation has shown that micro roller embossing has many advantages such as being suitable for large area ceramic patterning and easy in demolding. It is proved that micro roller embossing is a promising method for patterning large area ceramic green substrates.

### Acknowledgment

The authors would like to express acknowledgments to Dr. SH Ng, Mr. SH Ling, Ms. HP Maw, Ms. L Jin, Mr. YN Liang and Dr. HY Gan of SIMTech for their helpful discussion and cooperation. This work is supported by Agency for Science, Technology and Research (A*STAR) of Singapore for Singapore-Poland Cooperation.